\documentclass{article}

\usepackage{arxiv}

\usepackage[utf8]{inputenc} 
\usepackage[T1]{fontenc}    
\usepackage{hyperref}       
\usepackage{url}            
\usepackage{booktabs}       
\usepackage{amsfonts}       
\usepackage{nicefrac}       
\usepackage{microtype}      
\usepackage{float}
\usepackage{graphicx}
\usepackage{caption}
\usepackage{subcaption}
\usepackage{natbib}
\usepackage{doi}
\usepackage{siunitx}
\usepackage{mathtools}
\usepackage{natbib}
\usepackage{wrapfig}
\usepackage{multicol}
\usepackage[dvipsnames]{xcolor}
\usepackage{tikz}
\definecolor{ao(english)}{rgb}{0.0, 0.5, 0.0}

\usepackage{amsfonts}
\usepackage{amsmath} 
\usepackage{amssymb}

\usepackage{breqn}

\makeatletter
\newcommand*{\transpose}{%
  {\mathpalette\@transpose{}}%
}
\newcommand*{\@transpose}[2]{%
  \raisebox{\depth}{$\m@th#1\intercal$}%
}
\title{Deconvolutional double-difference misfit measurements and the application for full-waveform inversion}

\author{ {Fuqiang~Chen, Daniel Peter}\\
\\
	Physical Sciences and Engineering (PSE)\\
	King Abdullah University of Science and Technology\\
	Thuwal, Jeddah 23599 \\
	\\
	\texttt{fuqiang.chen@kaust.edu.sa} \\
}

\renewcommand{\shorttitle}{FWI with the deconvolutional double-difference misfit}

\graphicspath{{./Figs/}}
\begin{document}
\maketitle

\begin{abstract}
It is challenging for full-waveform inversion to determine geologically informative models from field data. An inaccurate wavelet can make it more complicated. We develop a novel misfit function, entitled deconvolutional double-difference misfit measurement to cancel the influence of wavelet inaccuracy on inversion results. Unlike the popular double-difference misfit measurement in which the first difference is evaluated by cross-correlation, the proposed one employs deconvolution to do this step. Numerical examples demonstrate that full-waveform inversion with the new misfit function is resilient to the wavelet inaccuracy. It can also converge to plausible local minima even from rough initial models.
\end{abstract}

\keywords{Deconvolutional double-difference \and Full-waveform inversion \and Seismic wavelet}

\section{Introduction}
Full waveform inversion (FWI) is now considered a useful method for retrieving knowledge of the earth's subsurface from observed seismic data \cite[]{VirieuxFWIoverview}. However, challenges of reaching its theoretical potential are always encountered as applied the method to field data. In this paper, we focus on one specific issue, which can generate a significant imprint of ambiguity in inversion results but only has been occasionally investigated, the source. For the sake of simplicity, the source mechanism is represented by the combination of a delta function and a wavelet in which the delta function and the wavelet indicate the source location and the time function stimulating the wave-propagation, respectively. In the case of active seismic, source locations are known. Therefore, a complete representation of the seismic source is only dependent on the wavelet. Because an exact recording cannot be obtained easily, the wavelet is then regarded as extra model parameters in the inversion process. It is estimated by either integrating into an extended model space with the subsurface parameters or more recently, FWI by source extension \cite[]{huang2019,symes2020,symes2021solution}, where a penalized misfit function of the wavelet is added to the data misfit. Both strategies require an accurate initial wavelet to prevent the inversion becoming more complicated to obtain geologically informative models. 

To remove the influence of wavelet inaccuracy on inversion results, this paper presents a novel misfit function based on the double-difference misfit measurement. The idea of double-difference misfit measurement can be dated back to \cite{poupinet1984} and \cite{Got1994}. Later on, \cite{Waldhauser2000} and \cite{Zhang2003} applied this idea for tomography to increase the mode resolution and improve the uncertainty of estimation. \cite{Duan2016} applied this idea to full wave-equation inversion. The double-difference misfit function minimizes the difference between two quantities induced from the observed and synthetic data, respectively. On the contrary, the conventional misfit functions directly minimize the difference between seismic datasets. We can consider the process of obtaining induced quantities as the first difference measurement. All applications of double-difference misfit measurement employ cross-correlation to perform this step. When applied to full wave-equation inversion \cite[]{Duan2016}, inverted results will be still impacted by the wavelet because cross-correlation is virtually limited to resist the outcome caused by the inaccuracy in wavelet, especially when inaccuracy means phase distortion. As we demonstrate later, the proposed double-difference misfit measurement can be immune to wavelet inaccuracy as the first differences are measured by deconvolution.
 
\section*{Deconvolutional double-difference misfit}
\subsection{Preliminaries}
According to the convolutional model assumption, a noise-free seismogram can be represented by the convolution of the earth's reflectivity with a seismic wavelet:
\begin{equation}
\mathbf{g}=\mathbf{r}\ast\mathbf{w},
\label{eq:cmodel}
\end{equation}  
where $\mathbf{g}$ is the synthesized seismic trace by the reflectivity $\mathbf{r}$ and seismic wavelet $\mathbf{w}$. We can generalize this theorem such that it is still formally valid for multi-dimensional velocity models.   
Suppose observed seismic traces $\mathbf{g}_1,$ $\mathbf{g}_2,$ $\cdots,$ $\mathbf{g}_{nr}$ are associated with a heterogeneous model. Then we may assume that those traces can be still represented by the convolutional model: 
\begin{subequations}
\begin{alignat}{5}
&\mathbf{g}_i &=& \mathbf{r}_i  &*&\mathbf{w},\\
&\mathbf{g}_{i+1} &=& \mathbf{r}_{i+1}  &*&\mathbf{w},
\end{alignat}
\label{eq:conv_model}
\end{subequations}
where $\mathbf{r}_i$ and $\mathbf{r}_{i+1}$ no longer mean the reflectivity; however, they are irrelevant with the wavelet $\mathbf{w}$. Then define a transfer function $\mathbf{d}_{i,0}$ connecting $\mathbf{g}_i$ and $\mathbf{g}_{i+1}$:
\begin{equation}
\mathbf{g}_i\ast\mathbf{d}_{i,0}=\mathbf{g}_{i+1}.
\label{eq:decon_obs}
\end{equation}
Substitute equation \ref{eq:conv_model} to equation \ref{eq:decon_obs}, we have 
\begin{equation}
\mathbf{r}_i\ast\mathbf{d}_{i,0}=\mathbf{r}_{i+1}.
\label{eq:decon_ref}
\end{equation}
Equation \ref{eq:decon_ref} informs us that $\mathbf{d}_{i,0}$ is independent of the wavelet. This quantity may be used to construct a misfit measurement free of wavelet effect. Even the original convolutional model is only valid for the 1D model, the example section demonstrates that the misfit function motivated by equation \ref{eq:decon_ref} can still remove the wavelet effect as applied to laterally heterogeneous model estimation.   
\subsection{Theory}
Let $\mathbf{f}_1,\mathbf{f}_2,\cdots,\mathbf{f}_{nr}$ stand for the predicted traces, we can also define a series deconvolution problem as 
\begin{equation}
\mathbf{f}_{i}\ast\mathbf{d}_{i,1}=\mathbf{f}_{i+1},
\label{eq:decon_syn}
\end{equation}
where $\mathbf{d}_{i,1}$ denotes the transfer function associated with $\mathbf{f}_i$ and $\mathbf{f}_{i+1}$. Then the deconvolutional double-difference misfit function can be defined as
\begin{equation}
\varepsilon=\displaystyle\sum_{i=1}^{nr-1}\mathbf{p}_i^\transpose\mathbf{p}_i,
\label{eq:ddd-misfit}
\end{equation}
where $\mathbf{p}_i=\mathbf{d}_{i,1}-\mathbf{d}_{i,0}$. 
  
Deconvolution problems in equations \ref{eq:decon_obs} and \ref{eq:decon_syn} can be reformulated into linear least-squares problems. Take equation \ref{eq:decon_syn} as an example, its matrix-vector form is
\begin{equation}
\mathbf{F}_{i}\mathbf{d}_{i,1}=\mathbf{f}_{i+1},
\label{eq:decon-mv-syn}
\end{equation}
where $\mathbf{F}_i$ represents the convolution matrix induced by $\mathbf{f}_i$.
Linear inverse models in equation \ref{eq:decon-mv-syn} are typically ill-posed as applied to seismic data. A well-known technique to solve this issue is Tikhonov regularization. The regularized linear inverse problems can be written as
\begin{equation}
\Big(\mathbf{F}^\transpose_{i}\mathbf{F}_{i}+\lambda^2\mathbf{D}^\transpose\mathbf{D}\Big)\mathbf{d}_{i,1}=\mathbf{F}_i^\transpose\mathbf{f}_{i+1},
\label{eq:decon-mv-syn-reg}
\end{equation}
where $\mathbf{D}$ can be the difference or identity matrix and $\lambda $ is the parameter to balance the smoothness of $\mathbf{d}_{i,1}$ and the minimization of mean errors in equation \ref{eq:decon-mv-syn}. 
The solution to equation \ref{eq:decon-mv-syn-reg} is then given by
\begin{equation}
\mathbf{d}_{i,1}=\Big(\mathbf{F}^\transpose_{i}\mathbf{F}_{i}+\lambda^2\mathbf{D}^\transpose\mathbf{D}\Big)^{-1}\mathbf{F}^\transpose_{i}\mathbf{f}_{i+1}.
\label{eq:aug-Axb}
\end{equation}

The adjoint source or the derivative of $\varepsilon$ w.r.t. $\mathbf{f}_i$ can be decomposed into two parts contributed by $\mathbf{d}_{i-1,1}$ and $\mathbf{d}_{i,1}$, respectively:
\begin{equation}
\frac{\partial \varepsilon}{\partial \mathbf{f}_i}=\Big(\frac{\partial \mathbf{d}_{i-1}}{\partial \mathbf{f}_i}\Big)^{\transpose}\frac{\partial\varepsilon}{\partial \mathbf{d}_{i-1}}+\Big(\frac{\partial \mathbf{d}_{i}}{\partial \mathbf{f}_i}\Big)^\transpose\frac{\partial\varepsilon}{\partial \mathbf{d}_{i}}.
\label{eq:depsilondf}
\end{equation}
According to equation \ref{eq:aug-Axb}, we directly obtain
\begin{equation}
\frac{\partial \mathbf{d}_{i-1}}{\partial \mathbf{f}_i}=\Big(\mathbf{F}^\transpose_{i-1}\mathbf{F}_{i-1}+\lambda^2\mathbf{D}^\transpose\mathbf{D}\Big)^{-1}\mathbf{F}^\transpose_{i-1},
\label{eq:sec_term}
\end{equation}
where we take the denominator-layout notations of matrix calculus.
On the contrary, ${\partial \mathbf{d}_i}/{\partial \mathbf{f}_i}$ cannot be directly obtained based on equation \ref{eq:aug-Axb} because $\mathbf{f}_i$ is not explicitly detached from the inverse term.
We define a three-dimensional matrix $\mathbf{H}$ such that
\begin{equation}
\mathbf{F}_i=\mathbf{H}\mathbf{f}_i,
\label{eq:mapping}
\end{equation}
where $\mathbf{H}\mathbf{f}_i$ converts $\mathbf{f}_i$ to a convolution matrix $\mathbf{F}_i$. 
Because $\mathbf{H}$ is a three-dimensional matrix, its transpose or permutation have $P_3^2=6$ possibilities. We use $\mathbf{H}^\transpose$ to represent the transpose from $\mathbf{H}=\mathbf{H}^{0,1,2}$ to $\mathbf{H}^{1,2,0}$ such that $\mathbf{f}^\transpose_{i}\mathbf{H}^\transpose$ corresponds to the transpose of the convolution matrix $\mathbf{F}_i$, where the superscripts ${0, 1, 2}$ denote the first, second, and last axis of matrix $\mathbf{H}$. Considering that convolution is interchangeable, we can have 
\begin{alignat}{5}
(\mathbf{Hx})\mathbf{y} &=& (\mathbf{Hy})\mathbf{x}.
\label{eq:Tprop}
\end{alignat}

Another useful property of $\mathbf{H}$ for the derivation of ${\partial \mathbf{d}_i}/{\partial \mathbf{f}_i}$ is 
\begin{equation}
(\mathbf{x}^\transpose\mathbf{H}^\transpose)\mathbf{y}=(\mathbf{H}^\transpose\mathbf{y})\mathbf{x}.
\label{eq:sec_prop}
\end{equation} 
Equation \ref{eq:sec_prop} states that the cross-correlation of $\mathbf{x}$ and $\mathbf{y}$ equals the cross-correlation of $\mathbf{y}$ and $\mathbf{x}$ in the reverse order.
Substitute equation \ref{eq:mapping} into equation \ref{eq:decon-mv-syn-reg}, we obtain
\begin{equation}
\Big(\mathbf{f}^\transpose_{i}\mathbf{H}^\transpose\mathbf{H}\mathbf{f}_{i}+\lambda^2\mathbf{D}^\transpose\mathbf{D}\Big)\mathbf{d}_{i,1}=\mathbf{f}_i^\transpose\mathbf{H}^\transpose\mathbf{f}_{i+1},
\label{eq:decon-reg-T}
\end{equation}
Apply the chain rule to implicit function $\mathbf{d}_{i,1}(\mathbf{f}_i)$ in equation \ref{eq:decon-reg-T}
and rearrange the derivative results based on properties in equations \ref{eq:Tprop} and \ref{eq:sec_prop}, we obtain 
\begin{equation}
\begin{split}
\frac{\partial \mathbf{d}_{i,1}}{\partial \mathbf{f}_i}=\Big(\mathbf{F}^\transpose_{i}\mathbf{F}_{i}+\lambda^2\mathbf{D}^\transpose\mathbf{D}\Big)^{-1}\Big(\mathbf{H}^\transpose\big(\mathbf{f}_{i+1}-\mathbf{F}_i\mathbf{d}_{i,1}\big)  -\mathbf{F}^\transpose_i\big(\mathbf{H}\mathbf{d}_{i,1}\big)\Big). 
\end{split}
\label{eq:adj_sec}
\end{equation}
With proper $\lambda$, the inverse terms from the aforementioned are always invertible. This makes the derivative of the proposed misfit measurements numerically accurate.
\begin{figure}
\centering
\begin{subfigure}{.4\textwidth}
  \centering
  \includegraphics[width=.90\linewidth]{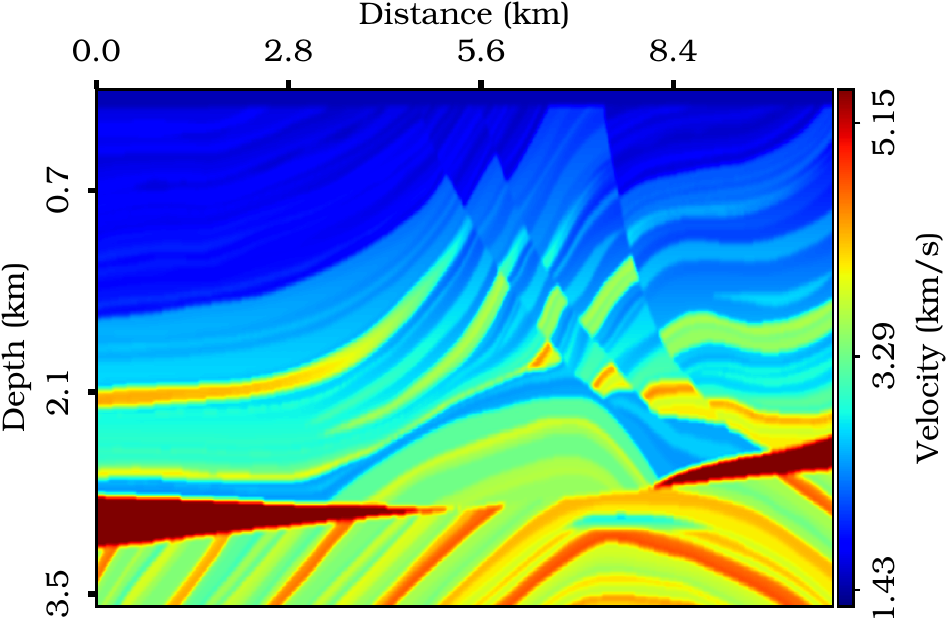}
  \caption{}
  \label{fig:true_marm}
\end{subfigure}%
\begin{subfigure}{.4\textwidth}
  \centering
  \includegraphics[width=.90\linewidth]{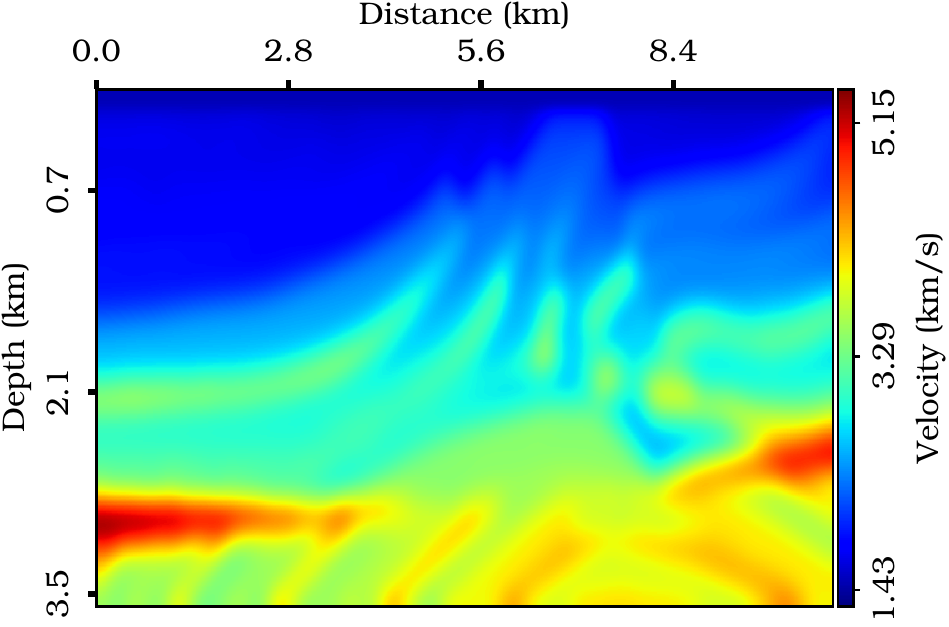}
  \caption{}
  \label{fig:init_marm0}
\end{subfigure}
\begin{subfigure}{.4\textwidth}
  \centering
  \includegraphics[width=.90\linewidth]{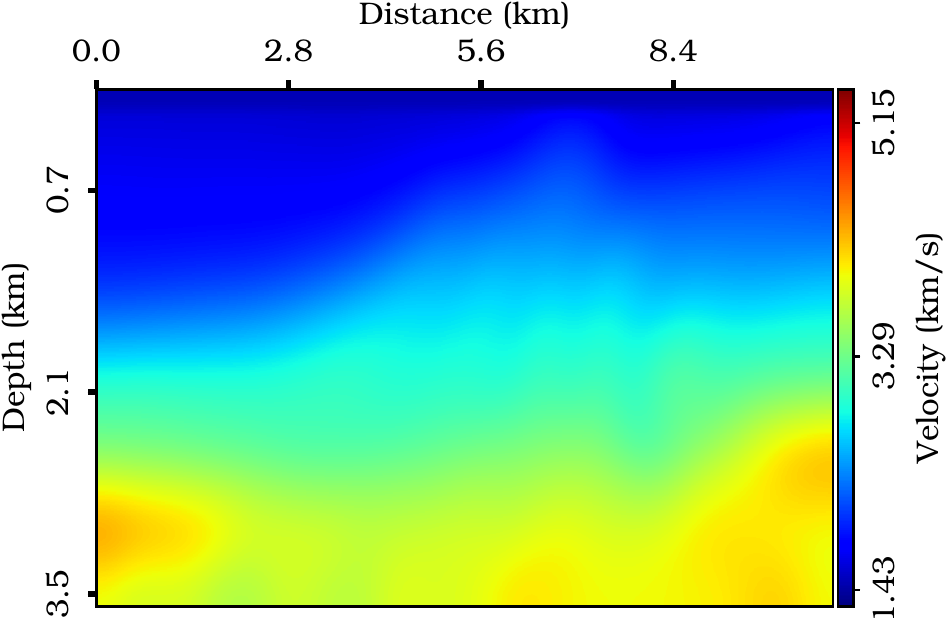}
  \caption{}
  \label{fig:init_marm1}
\end{subfigure}%
\begin{subfigure}{.4\textwidth}
  \centering
  \includegraphics[width=.90\linewidth]{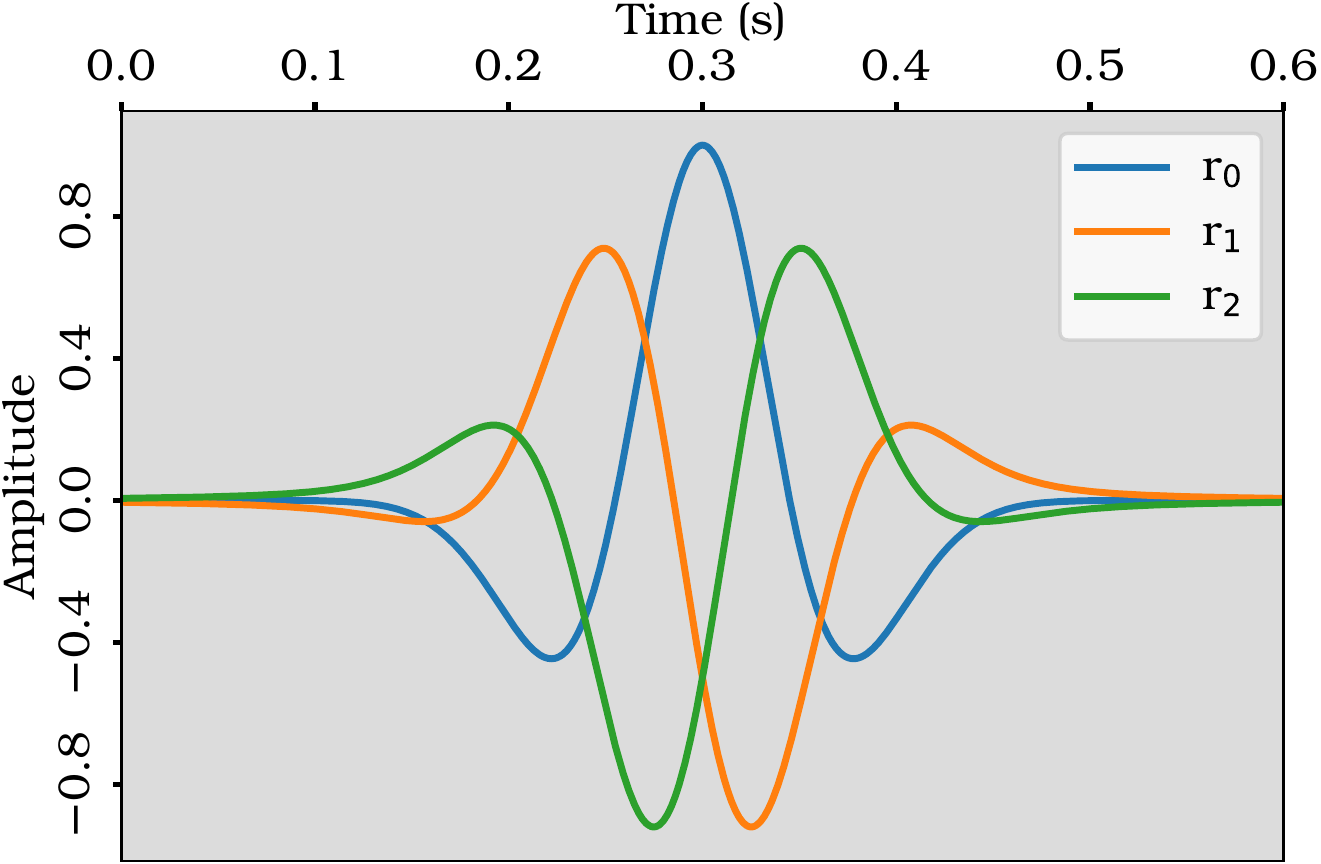}
  \caption{}
  \label{fig:wavs}
\end{subfigure}
\caption{(a) The true and (b and c) two initial models. (d) exact (blue) and inexact wavelets (orange and green).}
\label{fig:true_init_marms_wavs}
\end{figure}

\section{Marmousi synthetic example}

In this section, we demonstrate the ability of the deconvolutional double-difference misfit function to eliminate the wavelet inaccuracy.
Figure \ref{fig:true_marm} shows the true model to generate the observed data. Figures \ref{fig:init_marm0} and \ref{fig:init_marm1} illustrate two initial models in which the one in Figure \ref{fig:init_marm0} is assumed a good initial and the one in Figure \ref{fig:init_marm1} is rough. A $\SI{10}{\hertz}$ Ricker wavelet in Figure \ref{fig:wavs} (blue line) is used to generate the observed data. The orange and blue lines are two variants of the exact Ricker wavelet through rotating the phase by $\SI{120}{\degree}$ and $\SI{-120}{\degree}$, respectively. We name these three wavelets by $r_0$, $r_1$, $r_2$ in the following. 

\begin{figure}
\centering
\begin{subfigure}{.4\textwidth}
  \centering
  \includegraphics[width=.90\linewidth]{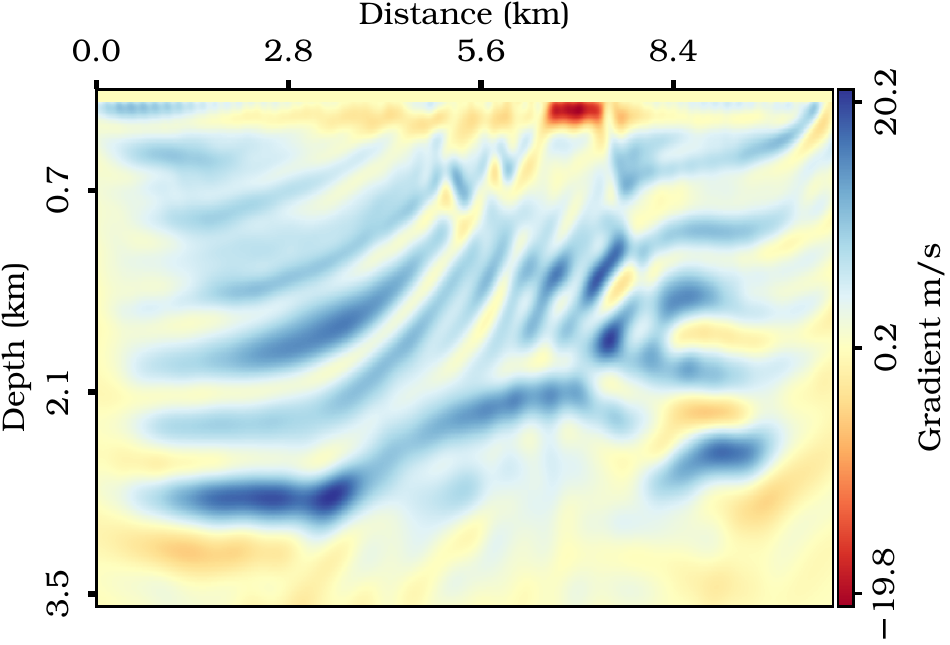}
  \caption{}
  \label{fig:marm_g_l2_00}
\end{subfigure}%
\begin{subfigure}{.4\textwidth}
  \centering
  \includegraphics[width=.90\linewidth]{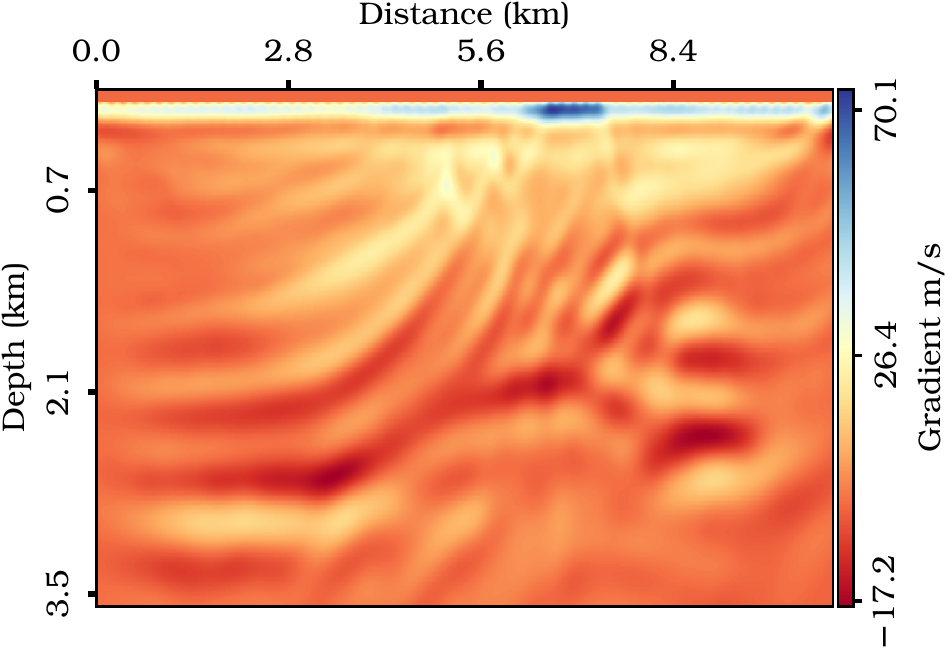}
  \caption{}
  \label{fig:marm_g_l2_120}
\end{subfigure}
\begin{subfigure}{.4\textwidth}
  \centering
  \includegraphics[width=.90\linewidth]{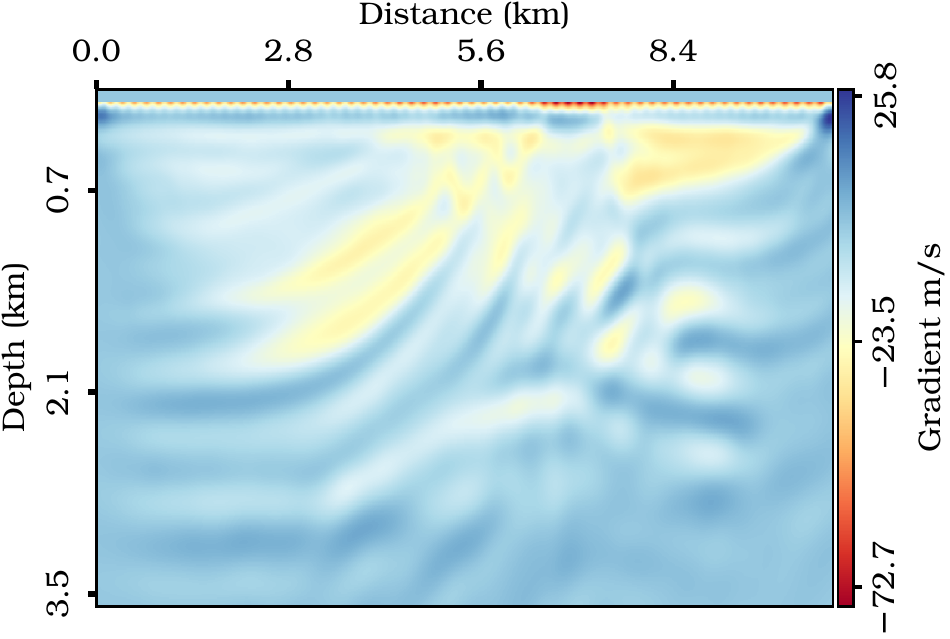}
  \caption{}
  \label{fig:marm_g_l2_240}
\end{subfigure}%
\begin{subfigure}{.4\textwidth}
  \centering
  \includegraphics[width=.90\linewidth]{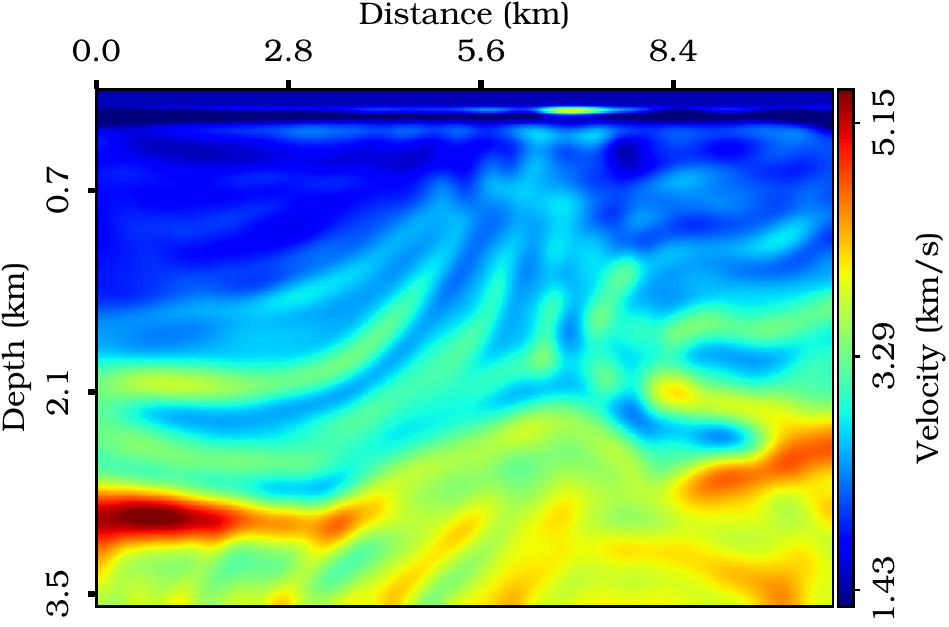}
  \caption{}
  \label{fig:marm_v_l2_0203}
\end{subfigure}
\caption{Gradient of misfit function based on $\mathcal{L}^2$ with (a) true, (b) $\SI{120}{\degree}$-rotated, and (c) $\SI{-120}{\degree}$-rotated wavelets, respectively. (d) The inverted model from data in $\SI{2}{\hertz}$--$\SI{3}{\hertz}$ by $\mathcal{L}^2$-norm misfit with the $\SI{-120}{\degree}$-rotated wavelet. }
\label{fig:marm_l2_g_v}%
\end{figure}

Figures \ref{fig:marm_g_l2_00} shows the gradient of misfit function based on $\mathbf{L}^2$-norm from the good initial model in Figure \ref{fig:init_marm0}, observed data in $\SI{2.0}{\hertz}$--$\SI{3.0}{\hertz}$, and the exact wavelet $\mathrm{r}_0$. Figures \ref{fig:marm_g_l2_120} and \ref{fig:marm_g_l2_240} show the same results but from inexact wavelets $\mathrm{r}_1$ and $\mathrm{r}_2$, respectively. By comparing the range of the gradient variation in Figures \ref{fig:marm_g_l2_00}--\ref{fig:marm_g_l2_240}, we can conclude that the wavelet inaccuracy can have a significant influence on the inversion result. Figure \ref{fig:marm_v_l2_0203} shows the inverted velocity with the inexact wavelet $\mathrm{r}_2$ after 20 iterations, where we see the shallow part approaches physically irrelevant model and it confirms that the wavelet inaccuracy causes a failure to converge to an expected output.      

\begin{figure}
\centering
\begin{subfigure}{.4\textwidth}
  \centering
  \includegraphics[width=.90\linewidth]{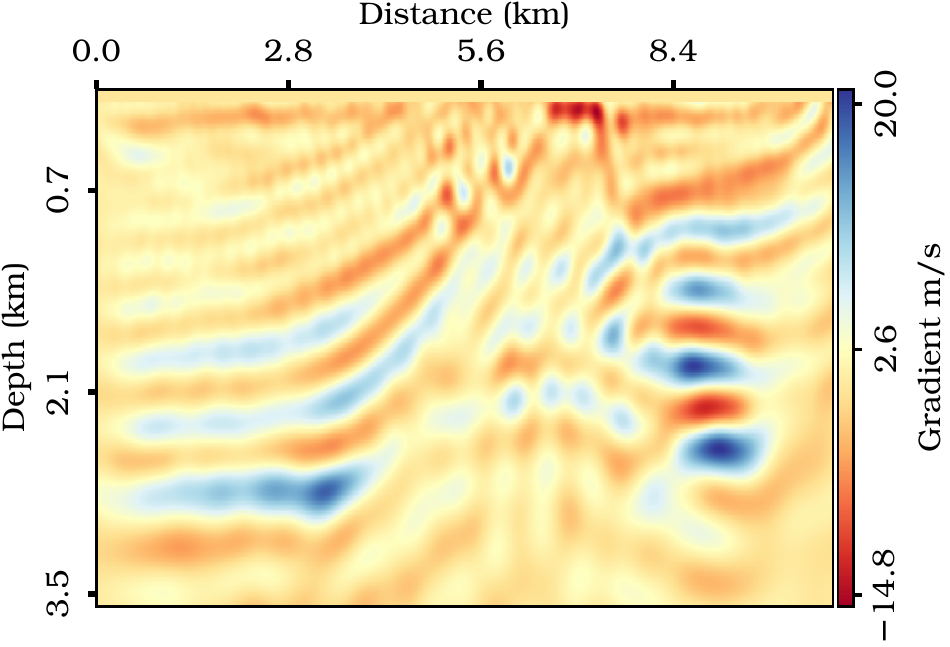}
  \caption{}
  \label{fig:marm_g_dd_00}
\end{subfigure}%
\begin{subfigure}{.4\textwidth}
  \centering
  \includegraphics[width=.90\linewidth]{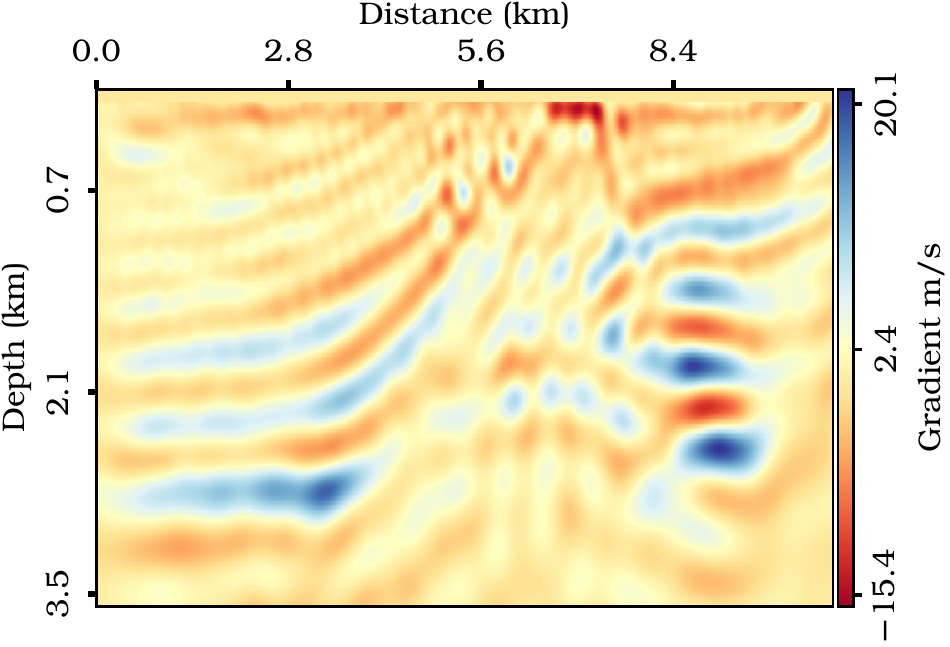}
  \caption{}
  \label{fig:marm_g_dd_120}
\end{subfigure}
\begin{subfigure}{.4\textwidth}
  \centering
  \includegraphics[width=.90\linewidth]{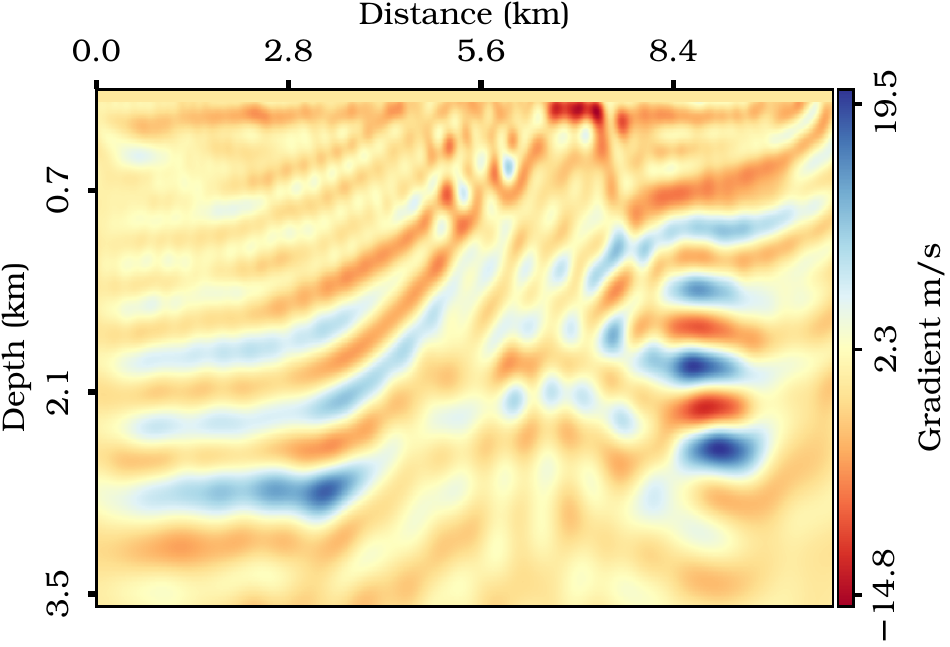}
  \caption{}
  \label{fig:marm_g_dd_240}
\end{subfigure}%
\begin{subfigure}{.4\textwidth}
  \centering
  \includegraphics[width=.90\linewidth]{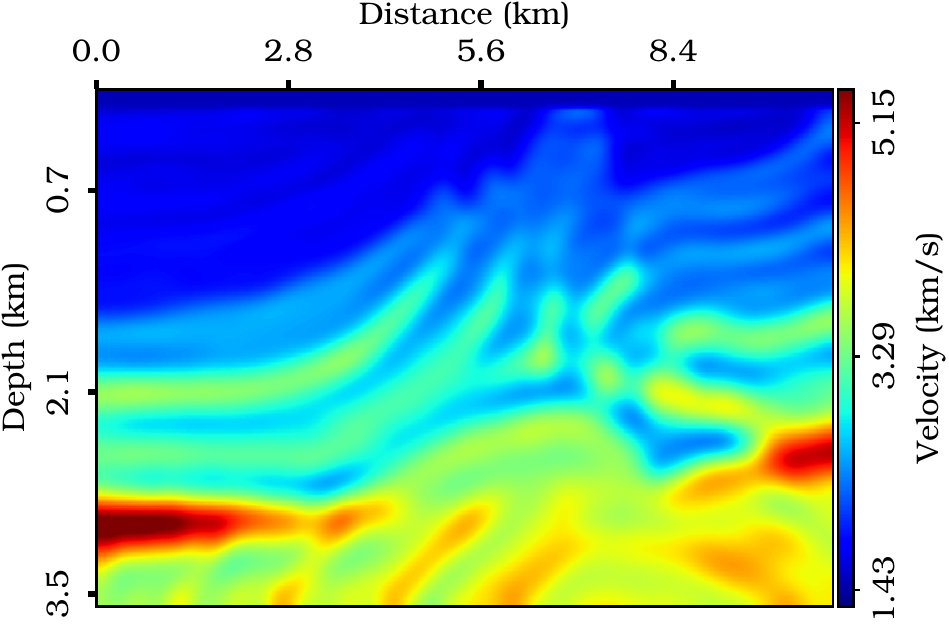}
  \caption{}
  \label{fig:marm_v_dd_0203}
\end{subfigure}
\caption{Gradient of the deconvolutional double-difference misfit function with (a) true, (b) $\SI{120}{\degree}$-rotated, and (c) $\SI{-120}{\degree}$-rotated wavelets, respectively. (d) The inverted model from data in $\SI{2}{\hertz}$--$\SI{3}{\hertz}$ by the proposed misfit with the $\SI{-120}{\degree}$-rotated wavelet. }
\label{fig:marm_dd_g_v}%
\end{figure}

Figure \ref{fig:marm_g_dd_00} shows the gradient of the proposed deconvolutional double-difference misfit function from the initial model in Figure \ref{fig:init_marm0}, observed data in $\SI{2.0}{\hertz}$--$\SI{3.0}{\hertz}$, and the exact wavelet $\mathrm{r}_0$. Figures \ref{fig:marm_g_dd_120} and \ref{fig:marm_g_dd_240} show the same results but from inexact wavelets $\mathrm{r}_1$ and $\mathrm{r}_2$, respectively. By comparing the range of the gradient variation in Figures \ref{fig:marm_g_dd_00}--\ref{fig:marm_g_dd_240}, we can conclude that the proposed misfit function can mostly mitigate the wavelet inaccuracy. Figure \ref{fig:marm_v_dd_0203} shows the inverted velocity with the inexact wavelet $\mathrm{r}_2$ after 20 iterations, where we can see that the proposed misfit measurement does not stagnate at irrelevant local minima with the inexact wavelet. 

\begin{figure}
\centering
\begin{subfigure}{.4\textwidth}
  \centering
  \includegraphics[width=.90\linewidth]{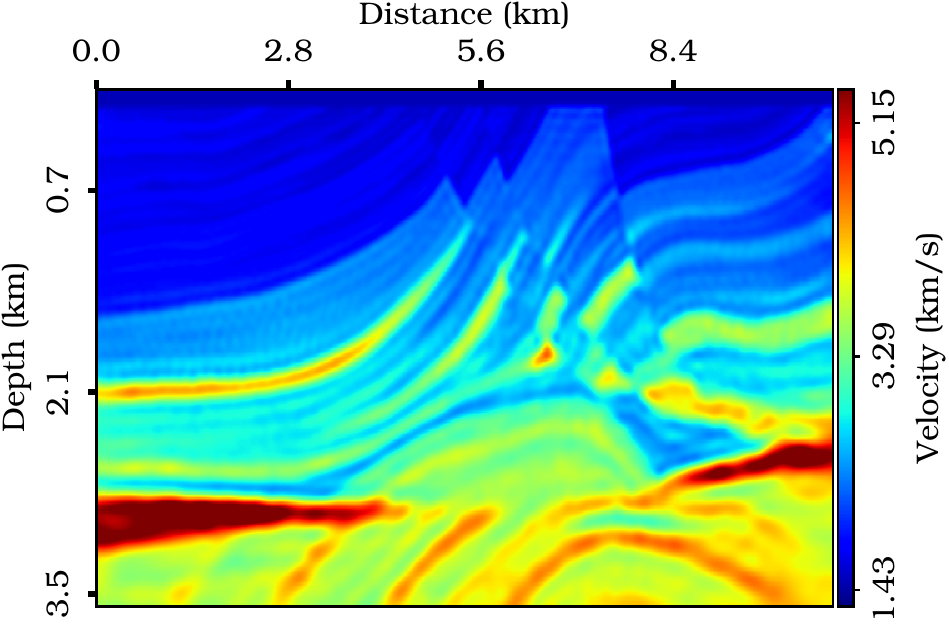}
  \caption{}
  \label{fig:marm_dd_v_10}
\end{subfigure}%
\begin{subfigure}{.4\textwidth}
  \centering
  \includegraphics[width=.90\linewidth]{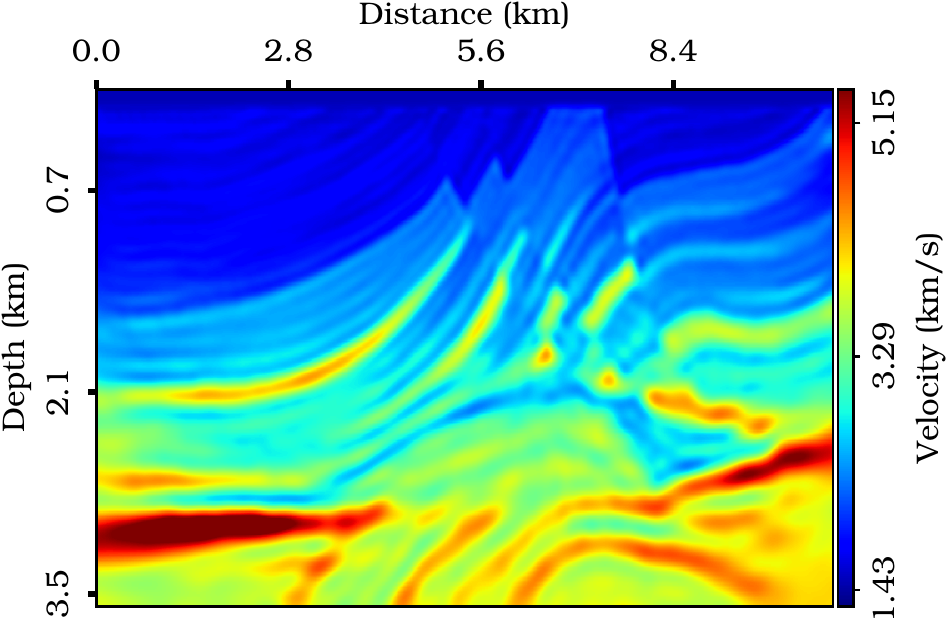}
  \caption{}
  \label{fig:marm_dd_v_8}
\end{subfigure}
\caption{Inverted model (a) from the initial model in Figure \ref{fig:init_marm0} and (b) from the one in Figure \ref{fig:init_marm1} by FWI with the proposed misfit function and the $\SI{-120}{\degree}$-rotated wavelet.}
\label{fig:marm_v_v}%
\end{figure}

Figure \ref{fig:marm_dd_v_10} shows the final estimated model from data in $\SI{2}{\hertz}$--$\SI{10}{\hertz}$ by frequency multi-scale strategy. Considering it starts with a relatively good initial model, we can see the proposed deconvolutional double-difference misfit function can converge to an informative model. Figure \ref{fig:marm_dd_v_8} shows the final estimated model from data in $\SI{2}{\hertz}$--$\SI{8}{\hertz}$ where the same recipes for the result in Figure \ref{fig:marm_dd_v_10} are adopted except the difference that a rough initial model in Figure \ref{fig:init_marm1} is used. We stop the inversion up to $\SI{8}{\hertz}$ because the decreasing in model misfit between the inverted and true model becomes insignificant.


\begin{figure}
\centering
\begin{subfigure}{.3\textwidth}
  \centering
  \includegraphics[width=.90\linewidth]{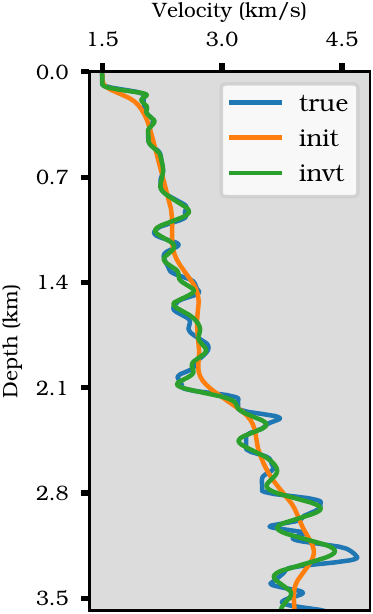}
  \caption{}
  \label{fig:wl-i0-0}
\end{subfigure}%
\begin{subfigure}{.3\textwidth}
  \centering
  \includegraphics[width=.90\linewidth]{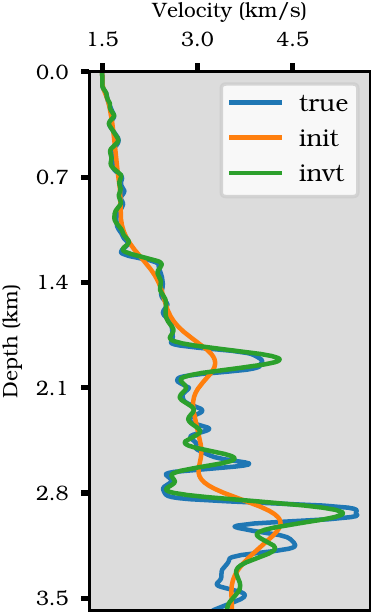}
  \caption{}
  \label{fig:wl-i0-1}
\end{subfigure}
\begin{subfigure}{.3\textwidth}
  \centering
  \includegraphics[width=.90\linewidth]{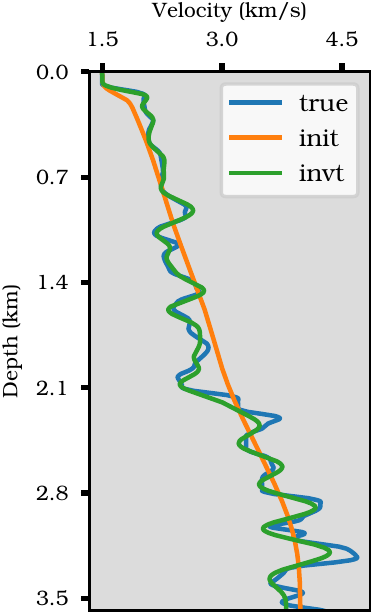}
  \caption{}
  \label{fig:wl-i1-0}
\end{subfigure}%
\begin{subfigure}{.3\textwidth}
  \centering
  \includegraphics[width=.90\linewidth]{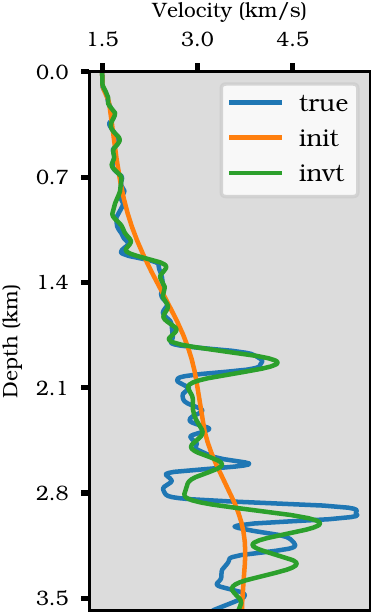}
  \caption{}
  \label{fig:wl-i1-1}
\end{subfigure}
\caption{The velocity profiles at $\mathbf{distance}=\SI{6.87}{\kilo\meter}$ (the first column) and $\mathbf{distance}=\SI{3.39}{\kilo\meter}$ (the second column). }
\label{fig:wl}%
\end{figure}

To evaluate the quality of inversion results from different initial models by the proposed misfit measurements, we extract the velocity profiles at two surface locations at $\mathbf{distance}=\SI{6.87}{\kilo\meter}$ (Figures \ref{fig:wl-i0-0} and \ref{fig:wl-i1-0}) and $\mathbf{distance}=\SI{3.39}{\kilo\meter}$ (Figures \ref{fig:wl-i0-1} and \ref{fig:wl-i1-1}). The orange and green lines in Figures \ref{fig:wl-i0-0} and \ref{fig:wl-i0-1} are extracted from the good initial in Figure \ref{fig:init_marm0}  and the inverted model in Figure \ref{fig:marm_dd_v_10}, where the well-fitting between inverted and true profiles illustrates that FWI with the proposed misfit function can provide informative information even the inversion starts with an inaccurate wavelet. Figures \ref{fig:wl-i1-0} and \ref{fig:wl-i1-1} further demonstrate the ability of the proposed misfit function to determine an informative estimate even from a rough initial model in Figure \ref{fig:init_marm1} and an inaccurate wavelet. 

\section{Conclusion}
We proposed employing deconvolution to evaluate the first difference measurement for double-difference misfit function. The new misfit function demonstrates its ability to remove the imprint in inversion results introduced by the wavelet inaccuracy. One concern about the double-difference misfit function is that it may converge to a physically irrelevant model because it does not measure the absolute difference and the low-frequency information will be missing after the first-difference measurement. However, numerical examples have shown that the proposed double-difference misfit function still can provide informative information even starting from a rough initial model.   

\section*{Acknowledgments}
We acknowledge the Supercomputing Laboratory at King Abdullah University of Science \& Technology (KAUST) for providing resources that contributed to the research results reported within this manuscript. 

\newpage
\bibliographystyle{seg}  
\bibliography{example}

\end{document}